\documentclass[11pt]{article}
\usepackage{epsfig,appb2,axodraw}

\newcommand{\sla}{\kern -5.4pt /}
\newcommand{\slaar}{\kern -7. pt / \kern 3.pt}
\newcommand{\Dir}{\kern -6.4pt\Big{/}}%su lettere italiane minuscole
\newcommand{\Dirin}{\kern -10.4pt\Big{/}\kern 4.4pt}
\newcommand{\DDir}{\kern -7.6pt\Big{/}}%su lettere italiane maiuscole
\newcommand{\DGir}{\kern -6.0pt\Big{/}}%su lettere greche

\newcommand{\ra}{\rightarrow}
\newcommand{\be}{\begin{equation}}
\newcommand{\ee}{\end{equation}}
\newcommand{\bea}{\begin{eqnarray}}
\newcommand{\eea}{\end{eqnarray}}
\newcommand{\beanon}{\begin{eqnarray*}}
\newcommand{\eeanon}{\end{eqnarray*}}
\newcommand{\ba}{\begin{array}}
\newcommand{\ea}{\end{array}}
\newcommand{\bd}{\begin{description}}
\newcommand{\ed}{\end{description}}
\newcommand{\bi}{\begin{itemize}}
\newcommand{\ei}{\end{itemize}}
\newcommand{\ben}{\begin{enumerate}}
\newcommand{\een}{\end{enumerate}}
\newcommand{\bc}{\begin{center}}
\newcommand{\ec}{\end{center}}

\newcommand{\ar}{\rightarrow}
\newcommand{\parno}{\par\noindent}

\def\pd{{\it production $\times$ decay\ }}
\def\epem{$e^+ e^-$\ }

\def\app #1 #2 #3 {{\it  Acta Phys.Polon.} {#1} (#2) #3\ }
\def\ap #1 #2 #3 {{\it Ann. Phys. }{ #1} (#2) #3\ }
\def\intj #1 #2 #3{{\it Int. J. Mod. Phys.} {#1} (#2) #3\ }
\def\hpa #1 #2 #3{{\it Helv. Phys. Acta. }{ #1} #2) #3\ }
\def\mpl #1 #2 #3 {{\it Mod.~Phys.~Lett.} {#1} (#2) #3\ }
\def\np #1 #2 #3 {{\it Nucl.~Phys.} {#1} (#2) #3\ }
\def\pl #1 #2 #3 {{\it Phys.~Lett.} {#1} (#2) #3\ }
\def\pr #1 #2 #3 {{\it Phys.~Rev.} {#1} (#2) #3\ }
\def\prep #1 #2 #3 {{\it Phys.~Rep.} {#1} (#2) #3\ }
\def\prl #1 #2 #3 {{\it Phys.~Rev.~Lett.} {#1} (#2) #3\ }
\def\rmp #1 #2 #3 {{\it Rev. Mod. Phys.} {#1} (#2) #3\ }
\def\zp #1 #2 #3 {{\it Z.~Phys.} {#1} (#2) #3\ }
\def\epj #1 #2 #3 {{\it Eur.~Phys.~J.} {#1} (#2) #3\ }
\def\cpc #1 #2 #3 {{\it Comp. Phys. Commun.} {#1} (#2) #3\ }
\def\xx #1 #2 #3 {{#1}, (#2) #3\ }

\begin{document}
\tolerance=100000
\thispagestyle{empty}
\setcounter{page}{0}

\begin{flushright}
{\large DFTT 34/98}\\
{\rm July 1998\hspace*{.5 truecm}}\\
\end{flushright}

\vspace*{\fill}

\bc
{\Large \bf 
Six Fermion Production at LC 
\footnote{ Talk presented at the 
Zeuthen Workshop on Elementary Particle Physics
{\it Loops and Legs in Gauge Theories},
Rheinsberg, 19--24 April 1998.\\[2 mm]
e-mail: ballestrero@to.infn.it}}\\[2.cm]

{\large  Alessandro Ballestrero }\\[.3 cm]
{\it I.N.F.N., Sezione di Torino, Italy}\\
{\it and}\\
{\it Dipartimento di Fisica Teorica, Universit\`a di Torino, Italy}\\
{\it v. Giuria 1, 10125 Torino, Italy.}\\
\ec

\vspace*{\fill}

\begin{abstract}
{\normalsize\noindent
We review  some features and results of the calculations performed 
with the program SIXPHACT for six fermion final states at Linear Collider.}
\end{abstract}
\vspace*{\fill}
\newpage

\section{Introduction}
An important part of LEP2 physics is concerned with 4-fermion final states.
From their studies\cite{yr}, it has become evident that on shell computations
 are often not sufficient for processes like for instance $WW$ production.
If on one hand {\it production $\times$ decay}
approximation makes it possible to compute various electroweak and strong
corrections, on the other one is neglecting
important issues such as irreducible backgrounds, finite width effects, spin 
correlations. The full set of diagrams for a given process is required for 
a complete description at parton level and to analyze distributions and 
experimental cuts.
The two approaches can be regarded at present as complementary, while 
corrections to the full set of diagrams start to be evaluated.
\par
With the advent of future \epem colliders, important physical processes,
such as $t\bar t$, $WWZ$ and Higgs production, will be concerned with 6-fermion 
final states. It is immediate for instance to realize that a final state like
that of $e^+e^-\ar \mu\bar\nu u \bar d b\bar b$ can be the result of 
the above mentioned production processes:

$e^+e^-\ar t\bar t\ar bW^+\bar b W^-\ar \mu\bar\nu u \bar d b \bar b$

$e^+e^-\ar W^+W^-Z\ar   \mu\bar\nu u \bar d b \bar b$

$e^+e^-\ar hZ\ar W^+W^-Z \ar \mu\bar\nu u \bar d b \bar b$
\parno
and of their irreducible backgrounds corresponding to all non resonant diagrams.
The same final state gets in reality many
different contributions which can be regarded as signal or background depending
on the physical process we are interested in.
Present helicity amplitude techniques allow to compute the tree-level full
set of diagrams for such final states in reasonable time.

Six fermion (6f) final states can be divided in charged, neutral and mixed
current processes:

{\bf CC} : 6f can form 2 W's and a Z but not  3 Z's 
               (e.g. $\mu\bar\nu u \bar d e^+e^-$)

{\bf NC} :  6f can form 3 Z's but not 2 W's and a Z
               (e.g. $u\bar u \mu^+\mu^- e^+ e^-$)

{\bf Mixed} :  6f can form both 3 Z's and 2 W's and a Z
               (e.g. $u\bar u d \bar d e^+e^-$)  
The program SIXPHACT \cite{noi} can compute all CC final states. It has been 
used to perform phenomenological studies of $t\bar t$, $WWZ$ and intermediate 
higgs physics. CC states are particularly interesting as they often allow
to exclude NC and QCD backgrounds. Specific processes have also been computed 
and analyzed by other groups. In particular the reaction 
$e^+e^-\ar\mu\bar \nu u \bar d b \bar b$ has been studied in
connection with $t\bar t$ production\cite{kuri} while
$e^+e^-\ar\mu^+ \mu^- l\bar\nu_l u \bar d$ $(l=e,\tau)$ and
$e^+e^-\ar e^+e^-\nu_e\bar\nu_e q\bar q$ ($q\neq b$)  have been examined 
for their relevance to intermediate mass higgs\cite{pv}.
\vfill\eject

\section{SIXPHACT. Method of calculation}
All diagrams for six fermion production can be subdivided into five different
topologies, corresponding to diagrams with quartic gauge coupling, two triple
gauge couplings, one triple gauge coupling, two fermion lines with two boson
insertions, one fermion line with three boson insertions. 
In spite of the fact that there are only few different topologies,
 the number of diagrams in CC 
processes is of the order of 200 when no exchanges among identical particles
are possible and it can raise to more than 1000 as it happens for $e^- \bar \nu
u \bar d e^+ e^-$ (1254). If we consider final states with one isolated lepton
 and four quarks,  $\mu \bar \nu u \bar d b \bar b$ has 232 diagrams and 
$e\bar \nu u \bar d u \bar u$ 840. 

The program for the amplitudes has been written with the help of 
PHACT \cite{phact}, a set of routines which implement the helicity method of 
ref.\cite{method}. With it, one can easily perform fast modular massive and
massless  amplitude computations. The key ingredients are so-called
$\tau$ matrices which assume a  trivial expression for fermion propagators.
The strategy used has been that of computing subdiagrams of increasing 
complexity, sum
them together when needed and compose the sums to form the final results.
In such a way repeated computations are avoided and the number of matrix
multiplications is optimized.

Taking as an example $e^+e^-\ar\mu\bar\nu u \bar d b \bar b$, we have first 
evaluated the subdiagrams corresponding to a boson in four fermions
or to the emission from the upper part of a fermion line of four outgoing or
two outgoing and two incoming fermions. If we indicate these subdiagrams with
the symbols:
\thicklines

%le misure sono in points. 1 point=.35mm. 
%quindi approssimativamente 3=1mm, 30=1cm. Cosi' il size della picture
%(che mi sembra circa il massimo possibile e' circa 14x21
\begin{picture}(420,70)(0,0)

%17
\SetOffset(0,30)
% gamma ff primario piccolo
\Photon(0,0)(30,0){2}{3}
\Text(10,-14)[lb]{$W$}
% le dimensioni di Oval sono i semiassi e per queste prima si da' y poi x     
\Oval(45,0)(18,15)(0)
\Text(37,12)[lt]{$u$ $\bar d$}
\Text(37,-12)[lb]{$b$ $\bar b$}

%16
\SetOffset(100,30)
% gamma ff primario piccolo
\Photon(0,0)(30,0){2}{3}
\Text(8,-14)[lb]{$\gamma Z$}
\Boxc(42,0)(24,30)
\Text(34,12)[lt]{$u$ $\bar d$}
\Text(34,-12)[lb]{$\mu$ $\bar \nu$}

%14
\SetOffset(220,30)
\Boxc(0,0)(24,30)
\Text(-8,12)[lt]{$u$ $\bar d$}
\Text(-8,-12)[lb]{$b$ $\bar b$}
\Line (0,15)(0,30)
\Line (0,-15)(0,-30)
\Text(-3,26)[rb]{$e^+$}
%\Text(-3,-34)[rt]{$1$}

%15
\SetOffset(300,30)
% le dimensioni di Oval sono i semiassi e per queste prima si da' y poi x 
\Oval(0,0)(18,15)(0)
\Text(-8,12)[lt]{$u$ $\bar d $}
\Text(-9,-12)[lb]{$e^+ e^-$}
\Line (0,18)(0,30)
\Line (0,-18)(0,-30)
\Text(-3,26)[rb]{$b$}
%\Text(-3,-34)[rt]{$1$}

\end {picture}
the computation of the full set of diagrams for 
$e^+e^-\ar\mu\bar\nu u \bar d b \bar b$ is then reduced to those in fig.~1.
\begin{figure}[h]
\vspace{0.1cm}
\begin{center}
\unitlength 1cm
\begin{picture}(10.,14.)
\put(-4.3,-2.){\includegraphics{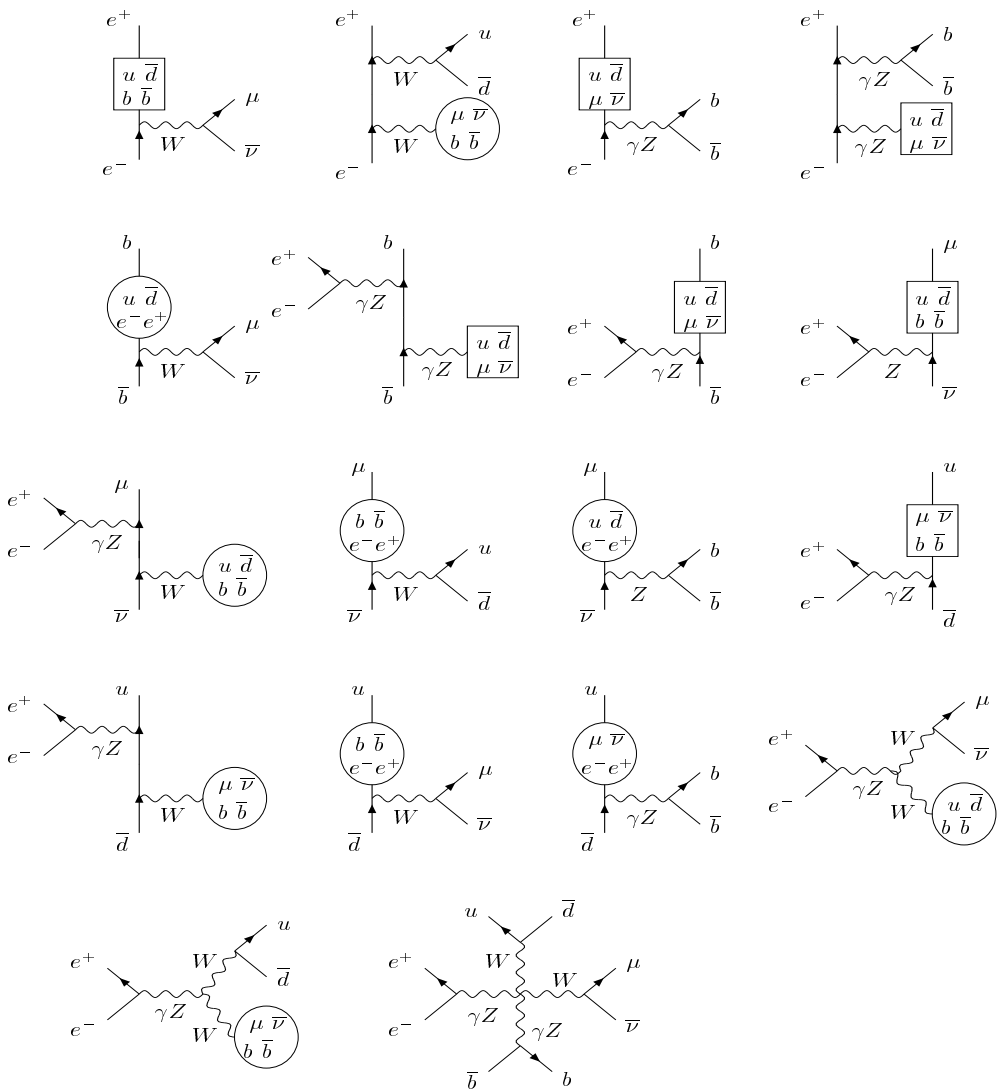}}
\end{picture}
\end{center}
\vspace{0.1cm}
\caption[]{Diagrams for the process  $e^+e^-\ar\mu\bar\nu u \bar d b \bar b$.
}
\label{f4}
\end{figure}

 SIXPHACT  can account for
 initial state radiation (ISR),
  beamstrahlung (BMS) with  a link to CIRCE\cite{circe},
  naive QCD corrections (exact in \pd no cuts limit),
  exact $b$ and $t$ fermion masses. All results given in the following have been
  computed in this way.

In order to account for the different peaking structures, SIXPHACT has 
several mappings for the 15-dimensional phase space.
It moreover automatically performs all requested distributions and any cut 
can be implemented. Apart from other specific cuts, we will 
require in the following jet(quark) energy $> 3$ GeV,
lepton energy $> 1$ GeV,
jet-jet invariant mass $> 10$ GeV,
lepton-beam angle $> 10^{\circ}$,
jet-beam angle $> 5^{\circ}$,
lepton-jet angle $> 5^{\circ}$. 

CPU times depend of course on the particular final processes, on the cuts,
and the options required. As an indication, a process like $e^+e^-\ar\mu^+
\mu^- e \bar\nu_e u \bar d$  with
ISR and BMS takes about 20 minutes for 3 per mille, 3 hours for per mille  and 
one day for .2 per mille accuracy on a DEC ALPHA station.

%\vfill\eject

\section{Applications}
Full six fermion calculations have been used to perform phenomenological 
studies of $t\bar t$ production in the continuum\cite{noi}\cite{kuri}, 
$WWZ$\cite{noi}  and 
intermediate mass higgs\cite{noi}\cite{pv}   physics 
at Linear Collider.

In this section we will limit ourselves to 
the last two issues, where for various final channels $t\bar t$ production
is to be considered as a huge background to be suppressed with appropriate cuts.
For brevity, we will not  attempt a complete discussion of the results but just
illustrate some examples.

\subsection{WWZ}
Triple gauge boson ($WWZ$) production processes are important as they 
allow a direct investigation of triple and quartic gauge boson couplings.
Accurate analyses of the sensitivities to genuine quartic coupling
have already been performed using only on shell calculation,  where the
three boson final state was considered\cite{wwz}. 
The cross sections are not very large: at $\sqrt{s}=500$ GeV 
 $WWZ$ production is of the order of 39 fb, which corresponds to about 
2000 events in all six fermion final states for an integrated luminosity
of 50 $fb^{-1}$. The announced possibility of having a final luminosity about 
ten times higher is therefore extremely important.

%
%\begin{table}[hbt]\centering
\begin{table}[h]\centering
\begin{tabular}{|c|c|c|c|}
\hline
%\rule [-0.25 cm]{0 cm}{0.75 cm}
process                         & WWZ NWA (fb) & WWZ signal (fb) & complete (fb)
\\
\hline \hline
%\rule [-0.25 cm]{0 cm}{0.75 cm}
$\mu \bar \nu u\bar d c\bar c$  & 0.13836(2)   & 0.13464(2)      & 0.16218(9) \\
%\cline{1-1} %\cline{4-4}
%\rule [-0.25 cm]{0 cm}{0.75 cm}
$e \bar \nu u\bar d c\bar c$    &            &             & 0.1783(2) \\
\hline
%\rule [-0.25 cm]{0 cm}{0.75 cm}
$\mu \bar \nu u\bar d s\bar s$  & 0.17780(3) & 0.17303(3)  & 0.1803(1) \\
%\cline{1-1} %\cline{4-4}
%\rule [-0.25 cm]{0 cm}{0.75 cm}
$e \bar \nu u\bar d s\bar s$    &            &             & 0.2117(2) \\
\hline
%\rule [-0.25 cm]{0 cm}{0.75 cm}
$\mu \bar \nu u\bar d u\bar u$  & 0.12815(2) & 0.12469(2)  & 0.1512(1) \\
%\cline{1-1} %\cline{4-4}
%\rule [-0.25 cm]{0 cm}{0.75 cm}
$e \bar \nu u\bar d u\bar u$    &            &             & 0.1758(3) \\
\hline
%\rule [-0.25 cm]{0 cm}{0.75 cm}
$\mu \bar \nu u\bar d d\bar d$  & 0.16468(3) & 0.16025(3)  & 0.16733(9) \\
%\cline{1-1} %\cline{4-4}
%\rule [-0.25 cm]{0 cm}{0.75 cm}
$e \bar \nu u\bar d d\bar d$    &            &             & 0.1941(1)  \\ 
\hline
\end{tabular}
\vskip .5cm
\caption[]{Cross section for the processes $e^+e^-\ra l\bar \nu_l + 4$ light
quarks ($l=\mu,e$) at $\sqrt{s}=500$ GeV}
\label{cswwz500}
\end{table}

We have considered final states with an isolated electron or muon and four 
final quarks: $e^+e^-\ar \mu (e)\bar\nu u\bar d q \bar q$. 
The isolated leptons give a signature that in the processes two $W$'s have been
produced, and are therefore useful in reducing the background.
We have also summed over all possible flavours for $q=u,d,c,s,b$. With no $b$
tagging one has the full contribution from $q=b$  which will be  
dominated by  $t\bar t$ production events. With a realistic $b$ tagging and 
appropriate cuts such background
can however be well under control\cite{noi}.
In table~1 we report the cross sections computed with the whole set of diagrams,
with only those diagrams corresponding to $WWZ$ \pd (signal)
and  the \pd approximation itself (Narrow Width Approximation).
The difference between full calculation and on shell results 
 is remarkable and shows the relevance of so called irreducible background.

\begin{figure}[h]
\vspace{0.1cm}
\begin{center}
\unitlength 1cm
\begin{picture}(10.,7.)
\put(-.3,-.8){\includegraphics{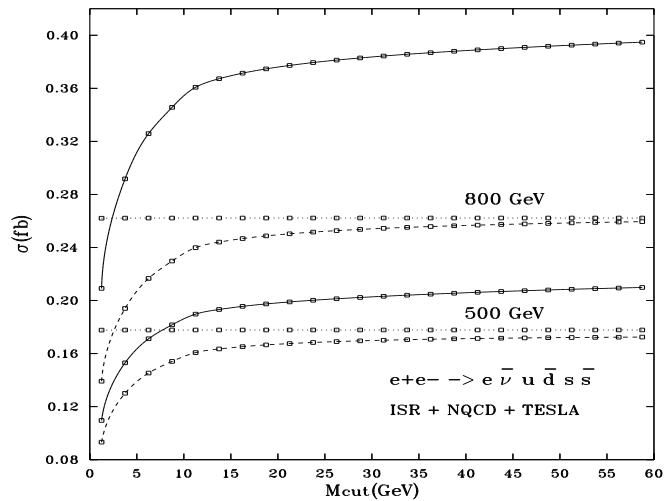}}
\end{picture}
\end{center}
\vspace{0.1cm}
\caption[]{Cross section for the process $e^+e^- \ar e^-\bar \nu_e u
\bar d s \bar s$ at $\sqrt{s}=500$ GeV (lower) and $\sqrt{s}=800$ GeV (upper)
as a function of $M_{cut}$.
Quarks are required to form two pairs whose invariant masses $m_i$ ($i=1,2$)
satisfy the conditions $|M_V -m_i| < M_{cut}$, $V=W,Z$.
The dot lines represent the cross section for $WWZ$ on shell, the dashed ones
the contribution of resonant $WWZ$ diagrams only, the continuous the complete
cross section. The markers indicate the points effectively computed.}
\label{f6}
\end{figure}

\begin{figure}[h]
\vspace{0.1cm}
\begin{center}
\unitlength 1cm
\begin{picture}(10.,7.3)
\put(-.3,-1.){\includegraphics{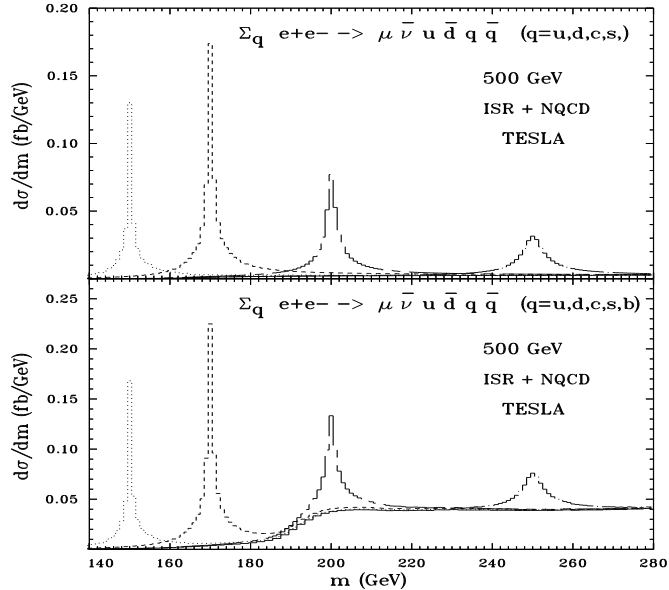}}
\end{picture}
\end{center}
\vspace{0.1cm}
\caption[]{{\it Reconstructed} mass distributions.
Quarks are required to form two pairs whose invariant masses $m_i$ ($i=1,2$)
satisfy the conditions $|M_V -m_i| < 20$ GeV, $V=W,Z$.
The continuous line represents the total 
background. The others correspond to the total cross sections for 
(from left to right) $m_h$= 150, 170, 200, 250 GeV.
 }
\label{h2}
\end{figure}

\begin{figure}[h]
\vspace{0.1cm}
\begin{center}
\unitlength 1cm
\begin{picture}(10.,7.3)
\put(-.3,-1.){\includegraphics{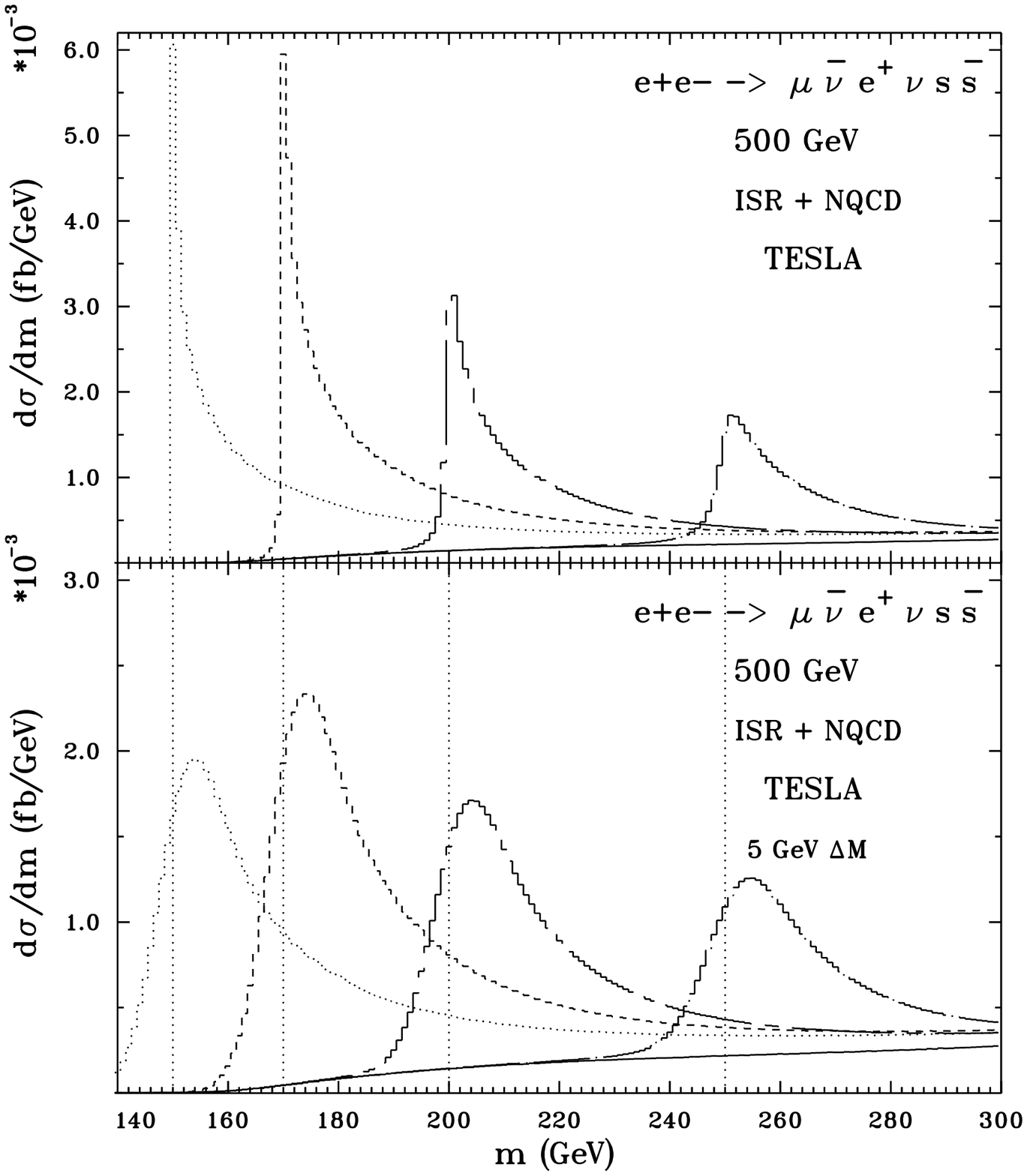}}
\end{picture}
\end{center}
\vspace{0.1cm}
\caption[]{{\it Missing} mass distributions. 
$s\bar s$ are required  to have an invariant mass $M$ such that
$|M_Z - M| < 20$ GeV. 
The continuous line represents the total 
background. The others correspond to the total cross sections for 
(from left to right) $m_h$= 150, 170, 200, 250 GeV.
In the lower plot a 5 GeV gaussian error in  missing mass determination is
assumed. 
}
\label{h4}
\end{figure}

In fig.~2  one uses cuts to force two of the invariant masses formed
by two quarks to be in the vicinity of the $W$ and $Z$ mass. Even with such 
a requirement, there is an evident difference between full calculation and 
resonant $WWZ$ diagrams. The effect grows with the energy. We have found
that it is less important for $e^+e^- \ar \mu \bar \nu_\mu u
\bar d s \bar s$, but in order to better isolate genuine $WWZ$ production,
cuts on the invariant mass formed by the lepton and the missing momentum
have to be studied.

\vfill\eject

\subsection{Intermediate mass Higgs}

If the mass of the Higgs is greater than about 140 GeV, it will mainly decay
to two $W$'s. In such a case, as the most important production
channels are  $hZ$ production (up to 500 GeV) and  $WW$ fusion (which dominates
at higher energies), the higgs events will effectively result in 6 fermion 
final states.

We have examined in detail  processes with one 
isolated lepton
like $l\: \nu_l\: +\: 4\: q's$ or $l\: \nu_l\: +\: l'\: \bar l'\: +\: 2\: q's$
and two leptons of different flavour and missing energy like 
$l\: \nu_l\: +\: l'\: \nu_{l'}\: +\: 2\: q's$. 
In the first two cases,
which represent respectively about 31\% and 4.4\% of $hZ$ signal, one can
study the distribution of the reconstructed mass, i.e. the invariant mass 
formed by
the isolated lepton, the reconstructed neutrino momentum and the two quarks
reconstructing the W. To the neutrino is attributed all  missing 3-momentum 
and its energy is taken to be equal to its modulus.
As far as $l\: \nu_l\: +\: l'\: \bar l'\: +\: 2\: q's$ is concerned, we have
verified that requiring that the invariant mass of the 2 $q$'s and of
$l'\: \bar l'$ be within 20 GeV from $m_W$ and $m_Z$ respectively, the
irreducible backgrounds become completely harmless. An analogous conclusion
can be drawn from the upper part of fig.~3 for the  case
of 4 light quarks and one isolated lepton. One would tend  to exclude
the $b$ quarks in such processes to get rid of the huge $t\bar t$ irreducible
background. In practice however $b$ flavour cannot be completely excluded with 
a realistic $b$ tagging and one would like to keep also the $b$ signal.
The lower part of fig.~3 shows that the cuts can still keep the background
under control, at least for not too high higgs masses, even if we sum also
over $q=b$ events.

In the case of $l\: \nu_l\: +\: l'\: \nu_{l'}\: +\: 2\: q's$ (about 5.2\% 
of $hZ$ signal) one cannot determine the reconstructed mass due to the presence
of the two neutrinos, but the missing mass can be used instead. This is
the invariant mass of the four-momentum recoiling against the particles 
decaying from the $Z$ (2 $q$'s in our case).
Fig.~4 shows that indeed one can very well see the signal also in such a case,
but the distribution around the higgs mass becomes asymmetric. In the more
realistic case in which a 5 GeV gaussian error in the determination of the 
missing mass is
assumed (lower part of fig.~4) one can see that the asymmetry implies a
consistent shift of the maximum with respect to the higgs mass.
 
\vfill\eject
\section{Conclusions}
Modern helicity techniques allow complete computation of many body final
states, which are of primary interest for future colliders. 
We have described some features of $e^+e^-\ar 6$ fermion calculations 
relevant to $t \bar t$, $WWZ$, intermediate higgs physics. 
The relevance of such an approach has been illustrated with some examples
of phenomenological results.

\section*{Acknowledgement}

I am grateful to Elena Accomando and Marco Pizzio for their
collaboration. I like to thank Tord Riemann and Johannes Bl\"{u}mlein  for the 
invitation and the  pleasant and constructive atmosphere during the 
workshop.

\vfill\eject
\end{document}